\documentclass[11pt]{article}
\usepackage{amssymb}
\usepackage{amsmath}
\usepackage{amstext}
\usepackage{graphicx,epsfig}
\usepackage{epsfig}
\usepackage{verbatim} 
\usepackage{fancybox}
\usepackage{color}
\usepackage{ulem}
\usepackage{enumitem}
\usepackage{subfigure}
\usepackage{bbm}
\usepackage{parskip}
\usepackage{dsfont}
\usepackage[numbers,sort&compress]{natbib}

\usepackage{hyperref} 	
\usepackage[all]{hypcap}  
\hypersetup{
    colorlinks=true,		
    linkcolor=red,		
    citecolor=blue,		
    filecolor=magenta,	
    urlcolor=black	
}

\newcommand{\be}{\begin{equation}}
\newcommand{\ee}{\end{equation}}
\newcommand{\bea}{\begin{eqnarray}}
\newcommand{\eea}{\end{eqnarray}}
\newcommand{\beas}{\begin{eqnarray*}}
\newcommand{\eeas}{\end{eqnarray*}}

\def\({\left(}
\def\){\right)}

\usepackage{myterms}

\newcommand{\QCD}{\text{\sc qcd}}

\newcommand{\EW}{\text{\sc ew}}

\newcommand{\QQQL}{\text{\sc qqql}}

\newcommand{\BpL}{\text{\sc b}+\text{\sc l}}

\newcommand{\tinyB}{\text{\sc b}}
\newcommand{\tinyL}{\text{\sc l}}
\newcommand{\tinyW}{\text{\sc w}}

\newcommand{\tinyZ}{\text{\sc z}}
\newcommand{\GammaB}{\Gamma_{\BpL}}
\newcommand{\thetaEW}{\bar{\theta}_{\EW}}

\title{}
\author{}

\numberwithin{equation}{section}

\begin{document}
%
~
\vspace{1truecm}
\renewcommand{\thefootnote}{\fnsymbol{footnote}}
\begin{center}
{\huge \bf{The Electroweak Vacuum Angle at Finite Temperature and Implications for Baryogenesis}}
\end{center} 

\vspace{1truecm}
\thispagestyle{empty}
\centerline{\Large Andrew J. Long,${}^{\rm a,}$\footnote{\tt andrewjlong@asu.edu} Hiren H. Patel,${}^{\rm b,}$\footnote{\tt hiren.patel@mpi-hd.mpg.de} and Mark Trodden${}^{\rm c,}$\footnote{\tt trodden@physics.upenn.edu}}
\vspace{.7cm}

\centerline{\it ${}^{\rm a}$Physics Department and School of Earth and Space Exploration,}
\centerline{\it Arizona State University, Tempe, Arizona 85287, USA.}

\vspace{.3cm}

\centerline{\it ${}^{\rm b}$Particle and Astro-Particle Physics Division, Max-Planck Institut fuer Kernphysik (MPIK)}
\vspace{.3cm}
\centerline{\it$^{\rm c}$Center for Particle Cosmology, Department of Physics and Astronomy,}
\centerline{\it University of Pennsylvania, Philadelphia, PA 19104, USA}

\vspace{.5cm}
\begin{abstract}
\vspace{.03cm}
\noindent
We initiate a study of cosmological implications of sphaleron-mediated CP-violation arising from the electroweak vacuum angle under the reasonable assumption that the semiclassical suppression is lifted at finite temperature.  
In this article, we explore the implications for existing scenarios of baryogenesis.  
Many compelling models of baryogenesis rely on electroweak sphalerons to relax a $(B+L)$ charge asymmetry.  
Depending on the sign of the CP-violating parameter, it is shown that the erasure of positive $(B+L)$ will proceed more or less quickly than the relaxation of negative $(B+L)$.  
This is a higher order effect in the kinetic equation for baryon number, which we derive here through order $n_{\BpL}^2$.  
Its impact on known baryogenesis models therefore seems minor, since phenomenologically $n_{\BpL}$ is much smaller than the entropy density.  
However, there remains an intriguing unexplored possibility that baryogenesis could be achieved with the vacuum angle alone providing the required CP-violation. 
\end{abstract}

\newpage


\newpage
\renewcommand*{\thefootnote}{\arabic{footnote}}
\setcounter{footnote}{0}

\section{Introduction}\label{sec:Introduction}

The origin of the baryon asymmetry of the universe, {i.e.} baryogenesis, remains one of the most compelling problems in cosmology and particle physics.  
A successful solution requires the violation of three global symmetries - baryon number (B), charge conjugation (C), and the composition of this symmetry with spatial parity (CP).  
A number of models take the minimal, and therefore attractive, approach of exploiting the anomaly-mediated baryon number violation due to electroweak sphalerons already present in the Standard Model (SM).  
As for CP-violation, however, the Standard Model is insufficient, and one has to rely on new physics.  
It has recently been pointed out~\cite{Perez:2014fja} that a CP-violating theta-term in the $\SU{2}$ electroweak (EW) sector, 
\begin{align}\label{eq:L_thetaEW}
	\Lcal_{\theta} = - \theta_{\EW} \frac{g^2}{32\pi^2} W_{\mu \nu}^{a} \widetilde{W}^{a \, \mu \nu} \com
\end{align}
can lead to CP-violation in the anomalous B-violating processes.  
In this paper, we initiate an exploration of how the electroweak vacuum angle $\theta_\EW$ may play a role in baryogenesis.  
For reviews of baryogenesis see \cite{Trodden:1998ym, Morrissey:2012db}.  

A number of phenomenological aspects of the electroweak vacuum angle have been studied in the literature~\cite{Anselm:1992yz, Anselm:1993uj, Lue:1998ke, McLerran:2012mm}, and these analyses have focused on zero-temperature observables, including scattering amplitudes and vacuum energy.  
In contrast, we investigate the effects of $\theta_{\EW}$-induced CP-violation at finite temperature and in the context of the early universe. 
At the outset it is important to recognize that this is a technical and difficult problem, since it entails quantum mechanical interference with non-perturbative processes at finite temperature.  
However, it is possible to make some progress without tackling these issues directly, and we are nevertheless able to explore the generic implications of the term~(\ref{eq:L_thetaEW}) for early universe cosmology, and to survey various potentially interesting avenues for future research.

In section~\ref{sec:EWAngle} we review the conditions under which the vacuum angle contributes to a physical phase $\thetaEW$, and we discuss the CP-violating observables associated with $\thetaEW$ at zero temperature.  
In section~\ref{sec:CPV_Finite} we study the implications of $\thetaEW$ for anomalous baryon number violation at finite temperature, and in section~\ref{sec:Baryogenesis} we explore the implications for baryogenesis.  
We conclude in section~\ref{sec:Conclusion} by summarizing our results and discussing directions for future work.

\section{Physics of The Electroweak Vacuum Angle}\label{sec:EWAngle}

Here we discuss zero-temperature observables associated with the electroweak vacuum angle, and we refer the interested reader to~\cite{Perez:2014fja} and~\cite{Anselm:1992yz, Anselm:1993uj} for further details.  

In the Standard Model, the electroweak vacuum angle is not physical, and the theta term, \eref{eq:L_thetaEW}, can be removed by an appropriate field redefinition, namely any combination of vector baryon-number (B) $\U{1}_{\tinyB}$ and lepton-number (L) $\U{1}_{\tinyL}$ transformations, 
\begin{equation}\label{eq:vecBLtransf}
	q \rightarrow e^{-i\alpha_{\tinyB}/3} q
	\qquad \text{and} \qquad
	\ell \rightarrow e^{-i\alpha_{\tinyL}} \ell \com
\end{equation}
satisfying $\alpha_{\tinyB}+\alpha_{\tinyL}=\theta_{\EW}/N_g$.  
Here $N_g=3$ is the number of fermion generations, and $q \in \{ Q_L, u_R, d_R \}$ and $\ell \in \{ L_L , e_R \}$ denote the quark and lepton fields, respectively.  
This conclusion can be avoided, however, in the presence of additional sources of $B$ and $L$ violation (either explicit or anomalous).  
For example, if we add to the lagrangian the dimension-6 operator 
\begin{align}\label{eq:L_QQQL}
	\Lcal_{\QQQL} = \lambda \frac{Q_L Q_L Q_L L_L}{\Lambda^2} \ ,
\end{align}
where $Q_L$ and $L_L$ denote the quark and lepton doublets, then the transformation~(\ref{eq:vecBLtransf}) no longer removes $\theta_{\EW}$ completely, but rather acts to transfer the electroweak vacuum angle onto the new operator:  $\Lcal_{\QQQL} \rightarrow e^{-i \theta_{\EW} / N_g} \Lcal_{\QQQL}$.  
Therefore, we may define an invariant phase 
\begin{align}\label{eq:thetaEW}
	\thetaEW = \arg\lambda + \frac{\theta_{\EW}}{N_g} 
\end{align}
that is unchanged under the field redefinition~(\ref{eq:vecBLtransf}).  
In this sense, the additional source of $B$ and $L$ violation supplied by the term~(\ref{eq:L_QQQL}) is analogous to the quark mass terms in QCD that render the QCD theta-term physical in the strong sector \cite{Callan:1976je}.  

The above argument is not unique to the operator~(\ref{eq:L_QQQL}), but instead a physical phase $\thetaEW$ emerges whenever the new operator mediates interactions with a $\Delta (B+L) \in \Zbb$ selection rule.  
For example, a much higher dimensional operator of the form $\frac{\lambda}{\Lambda^{14}}(Q_L Q_L Q_L L_L)^3$ would work.  
Another possibility occurs in left-right symmetric models, in which there exists a second ($B+L$) anomaly sourced by instantons in the $\SU{2}_\text{R}$ sector \cite{Anselm:1993uj}.  
In this paper, to keep the following discussion concrete, we will occasionally employ the interaction in \eref{eq:L_QQQL} as an explicit example, but our broader conclusions are not contingent upon this specific assumption, and in general the physical CP-violating phase $\thetaEW$ need not take the form as in \eref{eq:thetaEW}.  

\begin{figure}[t]
\begin{center}
\includegraphics[width=0.40\textwidth]{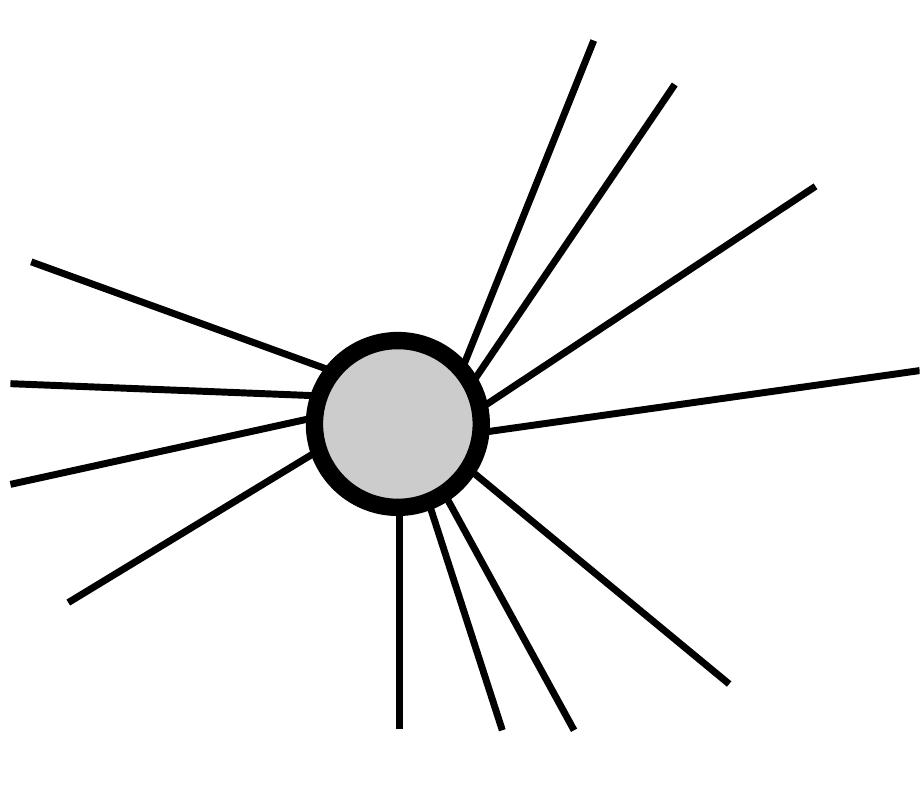} \qquad
\includegraphics[width=0.40\textwidth]{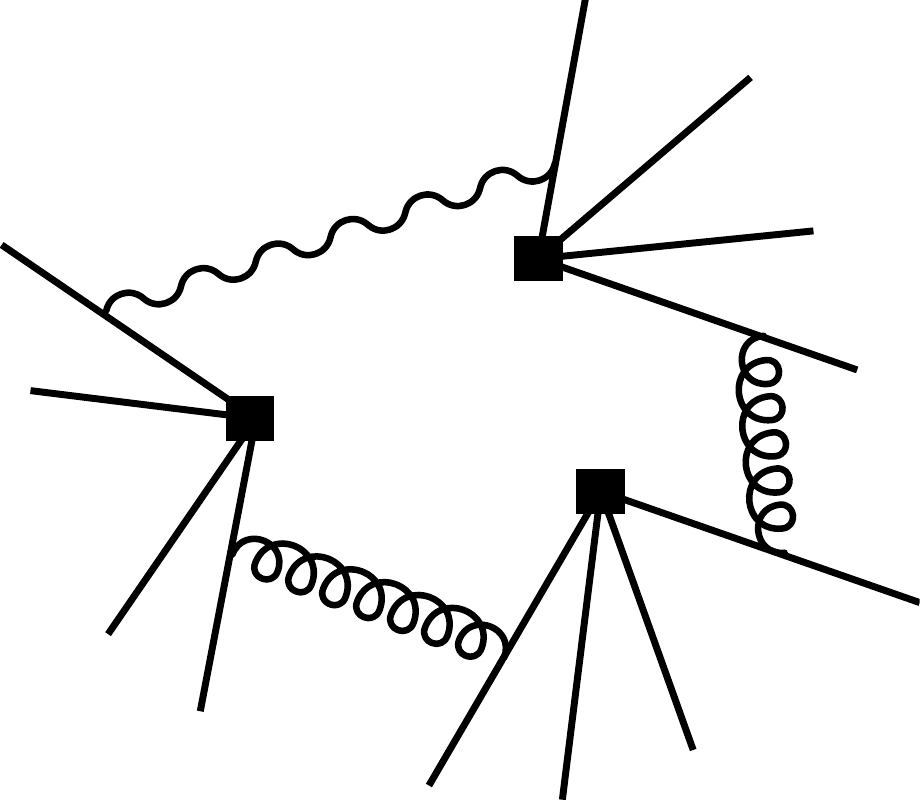} 
\caption{
\label{fig:interference}
CP-violation due to the electroweak vacuum angle arises as an interference between the EW instanton in the form of the effective 't Hooft vertex (left), and the new source of $B+L$ violation (right).  The square vertices are derived from \eref{eq:L_QQQL}.
}
\end{center}
\end{figure}

For $\thetaEW \neq 0$ the lagrangian is CP-violating and the phenomenological effects arise due to an interference between the electroweak instanton and the additional $B$ and $L$ violating interaction(s).  
The Feynman diagrams illustrating this point are shown in Fig. \ref{fig:interference}.  
A classic example exhibiting CP-violation is the following instanton-mediated $(B+L)$-violating process.  
Consider the pair of parton-level reactions related by CP conjugation~\cite{Ringwald:2002sw}
\begin{align}
\begin{array}{l}
\text{X}: q + q \rightarrow 3\bar\ell + 7\bar q\\
\overline{\text{X}}: \bar{q} + \bar{q} \rightarrow 3\ell + 7 q \ ,
\end{array} 
\end{align} 
where $q$ and $\ell$ stand for quarks and leptons.    
In the absence of CP-violation ($\thetaEW = 0$) the corresponding amplitudes, $\Mcal_{X}$ and $\Mcal_{\bar{X}}$, are identical.  
However, for $\thetaEW\neq 0$ we anticipate an induced asymmetry between the transition rates that can be quantified with the dimensionless parameter:
	\begin{align}\label{eq:epsilon_def}
	\varepsilon_{X} \equiv \frac{|\Mcal_{X}|^2 - |\Mcal_{\bar{X}}|^2}{|\Mcal_{X}|^2 + |\Mcal_{\bar{X}}|^2} \per
\end{align}
Since CP-violating effects derived from the electroweak vacuum angle involve the electroweak instanton (see \fref{fig:interference}), the asymmetry parameter $\varepsilon_{X}$ is suppressed by the usual semiclassical factor $\sim e^{-8\pi^2/g^2} \approx 10^{-80}$ (only one factor of $e^{-S_E}$ appears in the cross term).  
It is thus safe to say that CP-violation arising from the electroweak vacuum angle is too small an effect to measure at a TeV-scale hadron collider\footnote{However see also Refs.~\cite{Mattis:1991bj, Tinyakov:1992dr, Guida:1993qy} and more recently \rref{Tye:2015tva} for arguments to the contrary.  }.  
Similarly, other low energy observables such as atomic and nuclear electric dipole moments would be insensitive to the electroweak vacuum angle.  
In the above estimates we have evaluated $g(m_W) \simeq 0.65$ at the electroweak scale, since the $\SU{2}$ sector is in the Higgs phase and we are working with constrained instantons \cite{Vainshtein:1981wh}.  
It is illustrative to contrast with QCD, which is in the confinement phase where the strong coupling allows for unsuppressed CP-violation, manifested, for example, in electric dipole moments.  

\section{Effects of the Electroweak Vacuum Angle at Finite Temperature}\label{sec:CPV_Finite}

It is well-known that the instanton suppression is lifted in a thermal system in which electroweak Chern-Simons number evolves diffusively via electroweak sphalerons, allowing anomalous B-violation to become efficient~\cite{Kuzmin:1985mm}.  
In the Higgs phase at temperature $T \lesssim m_W$, the rate of baryon number violation is proportional to the Boltzmann factor $e^{-E_{\rm sph} / T}$ where $E_{\rm sph} \approx 2 m_W / \alpha_{\tinyW} \simeq 5 \TeV$ is the energy of the static sphaleron field configuration, $m_W$ is the W-boson mass, and $\alpha_{\tinyW} = g^2/4\pi$ is the electroweak fine structure constant \cite{Klinkhamer:1984di}.  
In the unbroken phase ($T \gtrsim m_W$), the energy barrier is absent and baryon number violation {via EW sphalerons occurs rapidly with a rate given by $\Gamma_{\rm sph} \approx \gamma_{\rm diff} / T^3$ where $\gamma_{\rm diff}$ is the Chern-Simons number diffusion coefficient (rate density) \cite{Khlebnikov:1988sr, Mottola:1990bz}.  }
The diffusion coefficient can be inferred from dimensional arguments~\cite{Arnold:1996dy} and lattice simulations~\cite{Bodeker:1999gx, DOnofrio:2012ni}, and one finds the rate of $(B+L)$-violation to be 
\begin{align}\label{eq:G_sph}
	\Gamma_{\rm sph} \simeq \kappa \, \alpha_{\tinyW}^{5} T 
\end{align}
with $\kappa \sim 100$.  Therefore at electroweak temperatures $\Gamma_{\rm sph}$ is much larger than the cosmological expansion rate $H \simeq 3 T^2 /  M_P$.  

In order to make robust quantitative statements regarding the role of $\thetaEW$ in early universe cosmology, we need to incorporate both the B-violating and CP-violating effects of the sphaleron in the transport equations.  
As mentioned in the previous section, the CP-violation can arise from an interference between perturbative operators and the instanton or sphaleron process (see Fig. \ref{fig:interference}).  
Independently, the two processes can be treated at finite temperature using well-established techniques: perturbative processes are conveniently computed using quantum mechanical matrix element methods, but the transition state theory computation of the Higgs phase sphaleron \cite{Affleck:1980ac}, or the evaluation of the sphaleron diffusion constant in the symmetric phase \cite{DOnofrio:2012ni} entirely bypasses the quantum amplitude, giving the rate directly.  
To our knowledge, a fully quantum mechanical treatment of the sphaleron process at finite temperature that retains the required phase information for interference is not known, and remains an important open problem.   

Despite this technical challenge, it is possible to make progress in understanding the CP-violating effects by asking how the electroweak vacuum angle arises in the kinetic equation for ($B+L$) number density.  
Writing the Noether current for $\U{1}_{\BpL}$ as $j_{\BpL}^{\mu}$, then the volume-averaged ($B+L$) charge density is given by 
\begin{align}
	n_{\BpL}(t) = \frac{1}{V} \int \! d^3x \, \langle j_{\BpL}^{0}(x) \rangle \per
\end{align}
As discussed in \aref{app:derive_G2}, the expectation value is calculated with respect to an out-of-equilibrium statistical ensemble that encodes the initial charge asymmetry~\cite{Khlebnikov:1988sr, Mottola:1990bz}.  
Using also the anomaly equation, 
\begin{equation}\label{eq:anomaly_equation}
	\partial_\mu j^\mu_{\BpL} = 2 \frac{N_g g^2}{32\pi^2}W_{\mu \nu}^{a} \widetilde{W}^{a \, \mu \nu}  \ ,
\end{equation}  
with $N_g = 3$ the number of fermion generations, one obtains the kinetic equation, 
\begin{align}\label{eq:kin_eqn_standard}
	\frac{dn_{\BpL}}{dt} = - \Gamma_{\rm sph} \, n_{\BpL}(t) \ ,
\end{align}
with the rate given by \eref{eq:G_sph}.  
An initial $(B+L)$ asymmetry is erased on a time scale $\Gamma_{\rm sph}^{-1}$, and the evolution is CP-symmetric, {i.e.} positive and negative charge asymmetries have identical evolutions.  

We now turn to the question of how these relaxation rates are modified by the presence of a non-vanishing electroweak vacuum angle.  
We assume new physics that contains $B$ and $L$ violating interactions, such as \eref{eq:L_QQQL}, and that there is a physical CP-violating phase $\thetaEW$.   
The new interaction provides a second channel through which $(B+L)$ erasure can be accomplished, and the rate in (\ref{eq:kin_eqn_standard}) then becomes $\Gamma_{\rm sph} \to \Gamma_{\BpL} = \Gamma_{\rm sph} + \Gamma_{\rm N.P.}$, where $\Gamma_{\rm N.P.}$ is the rate of $(B+L)$-violation induced by the new physics.  

\begin{figure}[t]
\hspace{0pt}
\vspace{-0in}
\begin{center}
\includegraphics[width=0.55\textwidth]{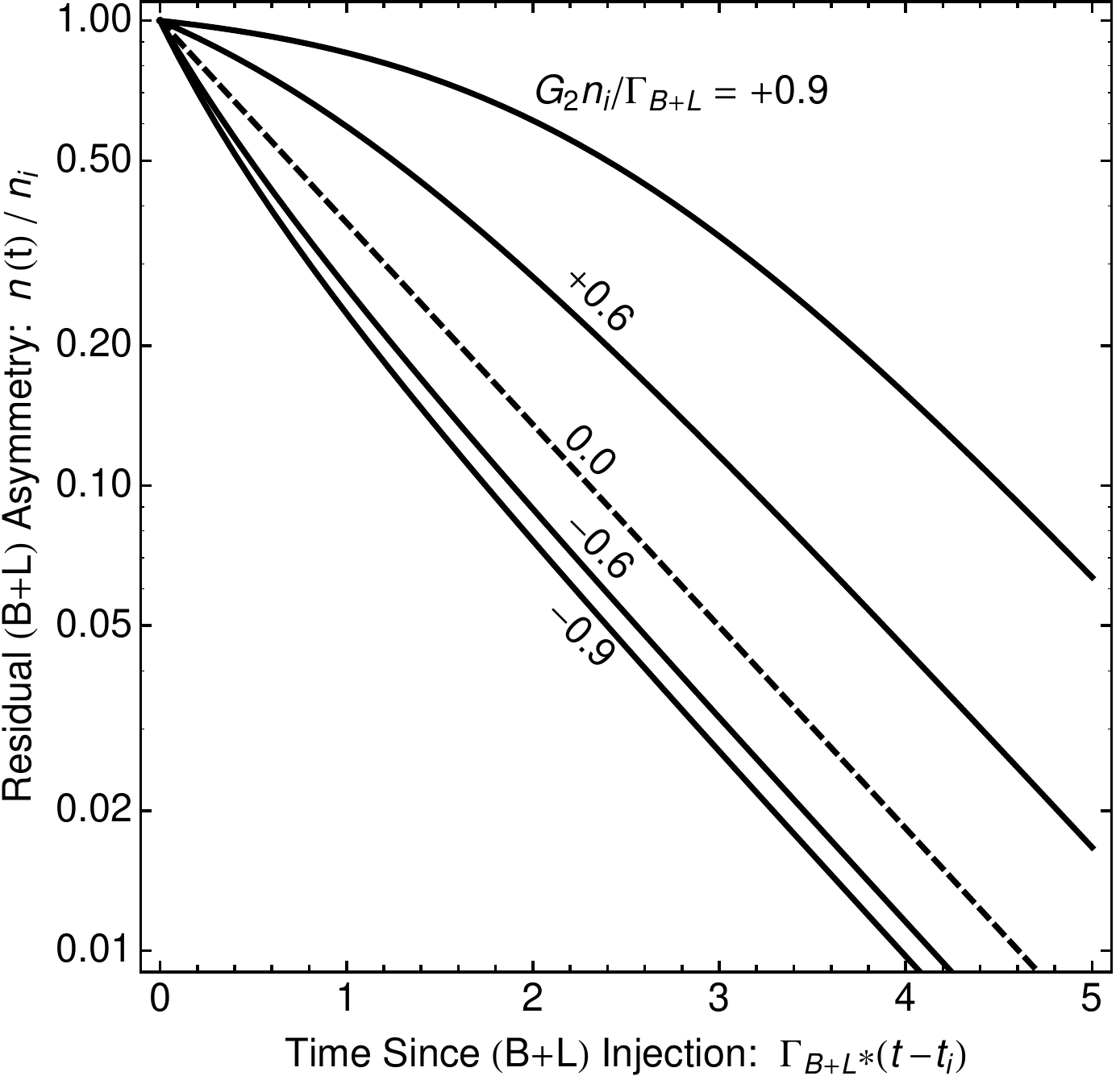} 
\caption{
\label{fig:solns}
The erasure of an initial $(B+L)$ asymmetry by electroweak sphalerons and explicit $(B+L)$-violating operators.  
The curves represent the solution (\ref{eq:solution}).  For $G_{2} = 0$ (dashed line) the evolution is CP-symmetric, {i.e.}, it is independent of the sign of the initial charge asymmetry.  For $G_{2} > 0$ an initial positive charge density will relax more slowly than a negative one, and for $G_{2} < 0$ the behavior is reversed.  
}
\end{center}
\end{figure}

However, because the rate equation (\ref{eq:kin_eqn_standard}) is invariant under CP reflection, $n_{\BpL} \to - n_{\BpL}$, this equation alone cannot describe CP-violation derived from the electroweak vacuum angle.  Rather, the CP-violating effects manifest themselves in terms higher order in $n_{\BpL}$.  To order $\mathcal{O}(n_{\BpL})^2$ the rate equation becomes
\begin{align}\label{eq:kin_eqn_extended}
	\frac{dn_{\BpL}}{dt} = - \Gamma_{\BpL} \, n_{\BpL}(t) + G_{2} \, n_{\BpL}(t)^2 + \cdots \ ,
\end{align}
where $G_2$ has  units of volume per time.  This form of the rate equation is no longer CP-invariant as long as $G_2 \neq 0$, and the corresponding solution then becomes
\begin{align}\label{eq:solution}
	n_{\BpL}(t) = \frac{n_{i} \, e^{- \Gamma_{\BpL} (t-t_i)}}{1 - \left( 1 - e^{- \Gamma_{\BpL} (t-t_i)} \right) \frac{G_2 n_{i}}{\Gamma_{\BpL}} } \per
\end{align}
The evolution described by this solution depends on a new dimensionless parameter $G_{2} n_{i} / \Gamma_{\BpL}$, which can be either positive or negative depending on the signs of $G_{2}$ and $n_{i}$.  
The solution is shown in \fref{fig:solns}.  
In the parameter regime $G_{2} n_{i} / \Gamma_{\BpL} \ll 1$ and the long-time limit $\Gamma_{\BpL}(t-t_i)\gg1$, the solution reduces to 
\begin{align}\label{eq:solution_2}
	n_{\BpL}(t) 
	& \sim \left( 1 + \frac{G_2 n_{i}}{\Gamma_{\BpL}} \right) \, n_{i} \, e^{- \Gamma_{\BpL} (t-t_i)} \per
\end{align}
For the case $G_2 n_{i} > 0$, the prefactor is larger than $1$ and the abundance is enhanced compared to the $G_{2} n_{i} < 0$ case.  
In this way we see how the parameter $G_{2}$ controls the effects of CP-violation.    

A formal derivation of (\ref{eq:kin_eqn_extended}) is presented in \aref{app:derive_G2}, where we express $G_{2}$ in terms of thermal expectation values.  
An explicit evaluation of $G_{2}$ is subject to the technical challenges described earlier, and is beyond the scope of this work. 
Instead, we perform a rough estimate of the dependence of $G_{2}$ on the parameters appearing in our model.  Recall that $G_{2}$ has the units of volume per time.  Since it parametrizes CP-violating effects, $G_{2}$ should be proportional to $\thetaEW$.  
Additionally, we expect a factor of $\gamma_{\rm diff} \sim \alpha_{\tinyW}^5 T^4$ associated with sphaleron diffusion in the electroweak unbroken phase, and three factors of $(\lambda / \Lambda^2)$ from the operator (\ref{eq:L_QQQL}) to achieve the required interference.  Combining these terms, and including additional factors of $T$ to obtain the correct mass dimension, we estimate $G_{2}$ as 
\begin{align}\label{eq:G2_est}
	 \sim \thetaEW \, \alpha_{\tinyW}^{5}T^4 \left(\frac{\lambda}{\Lambda^2}\right)^3 
\end{align}
up to a loop factor that may involve additional factors of the gauge couplings.  
The dimensionless quantity appearing in the solutions above is then estimated as 
\begin{align}\label{eq:G2_ratio}
	\frac{|G_{2}| n_{i}}{\Gamma_{\BpL}} \sim  \thetaEW \, \lambda^3 \frac{T^6}{\Lambda^6} \left( \frac{n_{i}}{n_{\gamma}} \right) \left( \frac{\Gamma_{\BpL}}{\alpha_{\tinyW}^5 T} \right)^{-1} \ ,
\end{align}
where we have used the photon number density, $n_{\gamma} \sim T^3$, as a normalization factor.  
In a successful baryogenesis model, the ratio $\eta \equiv n_{i} / n_{\gamma}$ will be at the level of the baryon asymmetry of the universe, $\eta \sim 10^{-10}$.   
Additionally, the validity of the effective field theory interpretation of \eref{eq:L_QQQL} requires $T < \Lambda$.  

This expression provides a useful estimate for the purposes of this paper, but our analysis of the implications for baryogenesis will not depend on this specific form for $G_{2}$.  However, the estimate does assume that the electroweak symmetry is restored and sphaleron diffusion occurs rapidly with a rate density $\gamma_{\rm diff} \sim \alpha_{\tinyW}^5 T^4$.  In the Higgs phase, sphaleron diffusion is Boltzmann suppressed, $\gamma_{\rm diff} \sim T^4 e^{- E_{\rm sph} / T}$, and we expect that $G_{2}$ acquires a similar suppression.  It is also worth noting that CP-violating effects in the sphaleron sector do not shift the equilibrium point away from $n_{\BpL} = 0$ \cite{Nauta:2000xi,Nauta:2002ru,Nauta:2003pt}.

It is possible that the electroweak vacuum angle may have other imprints at finite temperature apart from the modified kinetic equation (\ref{eq:kin_eqn_extended}), and this would be an interesting avenue to explore.

\section{Potential Implications for Baryogenesis}\label{sec:Baryogenesis}

Sphalerons play a central role in many of the most compelling models of baryogenesis, including electroweak baryogenesis and leptogenesis.  
In these models, perturbative particle physics yields a CP-asymmetry, and sphalerons then convert this into a baryon asymmetry as they equilibrate $(B+L)$ \cite{Kuzmin:1985mm, Shaposhnikov:1986jp}.  
Since the new physics we are discussing changes the process of charge relaxation by sphalerons, as seen in (\ref{eq:solution}), it is important to determine the impact this may have on existing models of baryogenesis, and to ask whether entirely new directions for baryogenesis appear.  

In this section we explore these possibilities. We will endeavor to keep our discussion general, and in particular we do not specify what new physics or even which operator has been added to the SM to yield the invariant phase $\thetaEW$.  Note that some operators such as $\Lcal_{\QQQL}$ in (\ref{eq:L_QQQL}) will violate $(B+L)$ explicitly, and one is inclined to ask whether baryogenesis can be accomplished without invoking the sphaleron.  A model of baryogenesis using the operator $\Lcal_{\QQQL}$ exists \cite{Chung:2001tp} in which the CP-violation is provided by relative phases between different elements of the flavor matrix $\lambda$ which are perturbative in origin.  Here we are considering a distinct source of CP-violation that can arise even when $\lambda$ is diagonal.  
Note that there is an analogous distinction in the Standard Model, where CP-violation due to $\delta_\text{CKM}$  arises from relative phases between elements of the quark Yukawa/mass matrix, but CP-violation arising from $\bar{\theta}_{\QCD}$ is a flavor-independent phase multiplying the overall matrix, and is manifestly non-perturbative in nature.

\subsection{Leptogenesis}\label{sec:Leptogenesis}

In the standard picture of leptogenesis \cite{Fukugita:1986hr} one extends the Standard Model to include a right-chiral Majorana neutrino with a mass well above the electroweak scale.  The Majorana neutrino decays into the SM Higgs and leptons, and if the interaction violates CP then there is a preference for decay into leptons over anti-leptons, or vice versa.  A lasting lepton asymmetry can be generated when these decays freeze out in the early universe at a temperature equal to the Majorana mass scale.  Above the electroweak scale sphalerons are in thermal equilibrium, and between freezeout and the electroweak phase transition they then convert the initial lepton asymmetry into the required baryon asymmetry. (For a review see \cite{Buchmuller:2004nz}.)  

In the models we are interested in, the conversion of a leptonic asymmetry to a baryonic one is governed by (\ref{eq:kin_eqn_extended}), and the challenge is to understand the effects of the CP-odd term with coefficient $G_{2}$ on the generation of the final baryon asymmetry.  As we have already seen in (\ref{eq:solution}), CP-violating effects are controlled by the dimensionless ratio $G_{2} n_{i} / \Gamma_{\BpL}$ where $n_{i} = n_{\tinyL}$ is the initial lepton asymmetry in the context of leptogenesis.  In order to generate a baryon asymmetry of the observed order of magnitude, $n_{\tinyB} \sim 10^{-10} n_{\gamma}$, the initial lepton asymmetry typically needs to be at the same level. Since (\ref{eq:G2_ratio}) implies that $G_{2} n_{i} / \Gamma_{\BpL} < 10^{-10}$, the CP-asymmetric evolution described by (\ref{eq:solution}) is probably too small to play a significant role in leptogenesis.  Moreover, since sphalerons remain in equilibrium for a long time, the asymptotic solution with $n_{\BpL} \approx 0$ is eventually reached, and any CP-asymmetry in the preceding evolution is erased.  

\subsection{Nonlocal Electroweak Baryogenesis}\label{sec:NonLocal}

Models of baryogenesis at the electroweak scale typically require the electroweak phase transition to be first order so that it proceeds through the nucleation and percolation of bubbles of broken electroweak symmetry.
These models are classified as local if both CP- and B-violation occur at the bubble wall, or nonlocal if B-violation occurs in front of the wall.  

In models of nonlocal electroweak baryogenesis \cite{Kuzmin:1985mm, Cohen:1990py, Cohen:1990it, Cohen:1993nk} particles scattering from the bubble wall experience a CP-violating interaction that injects a CP-asymmetry into the plasma ahead of the wall.  As this asymmetry diffuses in front of the bubble wall,  sphalerons then act to process it into a baryon asymmetry via (\ref{eq:kin_eqn_extended}).  Finally, the bubble wall passes, and the baryon asymmetry is then in the interior of the bubble, where baryon number violation is negligible, and the newly created baryon number is thus preserved.  

When investigating the effect of CP-violation in the sphaleron sector on electroweak baryogenesis, we encounter the same challenge that arose in the case of leptogenesis; namely, the CP-asymmetry created at the bubble wall is already very small, and the higher order CP-violating effects in (\ref{eq:kin_eqn_extended}) do not play any significant role.  Partly because of this, it is interesting to consider the possibility that the electroweak vacuum angle might provide all of the requisite CP violation to achieve baryogenesis at the electroweak scale and thereby obviate the need for CP-violating interactions in the Higgs sector.  Such a scenario would necessarily go beyond the discussion of \sref{sec:CPV_Finite} in which CP-violation arises through the way that sphalerons process an initial CP-asymmetry.  Instead, here we would be imagining that sphaleron-mediated transitions {\it alone} could give rise to both the CP- and B-asymmetries.  

One possibility is that electroweak sphalerons at the bubble wall might induce the appropriate CP- and B-violating scatterings of particles in the plasma, as in the example in \sref{sec:EWAngle}, such that a net baryon number is created and diffuses into the bubble, where sphalerons are out of equilibrium.  
In this way, sphalerons at the bubble wall would provide the source of B-number.  This scenario has its own challenges.  
Within the bubble wall the Higgs condensate is nonzero, $0 < \langle \Phi^0 \rangle < v$, and the rate of sphaleron-mediated reactions acquires a Boltzmann suppression $e^{-E_{\rm sph}/T}$ with $E_{\rm sph} \approx 4 \pi \langle \Phi^{0} \rangle / g \sim 5 \TeV$.  
Since sphalerons are sparse in the wall, it seems unlikely that the scattering of particles could be efficient enough to generate the observed baryon asymmetry.  
Moreover, the vacuum angle alone may not be sufficient to bias the creation of a global B-asymmetry, even though it allows for B-violation in individual scattering processes.  
In other words, unitarity of the S-matrix and CPT invariance might forbid the generation of a global asymmetry via inelastic scattering processes alone \cite{Baldes:2014gca,Baldes:2014rda}.  

\subsection{Local Electroweak Baryogenesis}\label{sec:Local}

In local models of electroweak baryogenesis, both CP and baryon number are violated at the same point in space, typically at the bubble wall, with a variety of different implementations in the literature (see, e.g.~\cite{Trodden:1998ym}).  In one such scenario \cite{Turok:1990in, Turok:1990zg, Dine:1990fj, Dine:1991ck}, the passage of the bubble wall leaves behind nontrivial field configurations with a net winding number that can be associated, in a specific gauge, with Chern-Simons number of the $\SU{2}$ gauge fields.  As these unstable configurations unwind to the vacuum, the associated level crossing can generate baryons and antibaryons through the anomaly, \eref{eq:anomaly_equation}.  
The inclusion of sufficient CP-violation, with an appropriate sign, then biases the unwinding of these configurations in favor of the production of a net baryon number.  

In a second scenario, the passage of the bubble wall ``kicks'' a sphaleron-like field configuration toward one of the adjacent vacua \cite{Shaposhnikov:1987pf, Dine:1990fj, Dine:1991ck}.  
Once again, appropriately chosen CP-violation biases kicks that increase Chern-Simons number, and a baryon asymmetry is generated by the associated level crossing.  

Beyond the framework of local electroweak baryogenesis, there exists a closely related class of non-thermal electroweak baryogenesis models.  
In these scenarios the universe is never heated above the electroweak scale and the electroweak phase transition does not occur.  
Nevertheless, nontrivial field configurations with winding can be produced nonthermally, for instance during the preheating epoch at the end of inflation or through a rapid quench \cite{Krauss:1999ng, GarciaBellido:1999sv, Enqvist:2010fd}, and CP-violation then biases the unwindings toward the generation of a baryon asymmetry.  

The key ingredient in these models is the incorporation of an appropriate source of CP-violation. One frequently considered possibility is to use the pseudoscalar operator $\Lcal = (b/M^2) \Phi^{\dagger} \Phi \, W \widetilde{W}$, which couples the Higgs field to the electroweak gauge fields.  
Unlike the theta term in \eref{eq:L_thetaEW}, this operator is not a total derivative; it directly affects the equations of motion, thereby leading to the necessary CP-asymmetric evolution.  
In contrast, the theta term can lead to an asymmetric evolution only as the result of a quantum interference \cite{Lue:1998ke, Hsu:2010jm,Hsu:2011sx}.  
For instance, an unwinding field configuration may generate a baryon number via level crossing, or it may not if there is interference with an explicit $(B+L)$-violating interaction, such as \eref{eq:L_QQQL}.  
In some sense, the anomaly creates $(B+L)$, and the perturbative operator immediately destroys it.  
It is interesting to consider whether $\thetaEW \neq 0$ provides a new source of CP-violation that might alter the usual predictions of nonlocal electroweak baryogenesis.

\subsection{Baryogenesis from a Large CP-Asymmetry}\label{sec:LargeAsym}

In the cases of leptogenesis and nonlocal electroweak baryogenesis, we have seen that the CP-violating effects of the vacuum angle have negligible impact on the final baryon asymmetry due to the smallness of the charge asymmetries compared to the entropy density. 
This is not due to the smallness of $\thetaEW$, and in fact this parameter is totally unconstrained.  
Instead it is a consequence of the small initial CP-asymmetry that the sphalerons act upon.  
It is interesting to consider how the CP-violating sphaleron transitions would affect baryogenesis if the CP-asymmetry were much larger.  
 
In the context of the electroweak phase transition, we might imagine that some mechanism generates a large, random CP-asymmetry in front of the bubble walls.  
This differs from usual nonlocal electroweak baryogenesis where each bubble produces a CP-asymmetry with the same sign.  
In a two Higgs doublet model with spontaneous CP-violation, for instance, different bubbles are formed from different linear combinations of the Higgs fields, and CP-violating interactions with the fermions can lead to CP-asymmetries of different signs in different regions of space.  

Due to the CP-violating effects discussed in \sref{sec:CPV_Finite}, sphalerons in front of the walls will process the positive and negative CP-asymmetries at different rates.  
Suppose that regions with $n_{\BpL} < 0$ relax back to zero net baryon number more quickly than regions with $n_{\BpL} > 0$.  
If sphalerons remain in equilibrium long enough, eventually the entire system will be bought back to the equilibrium point at which $n_{\BpL} = 0$.  
However, if the phase transition ends before the relaxation is complete, there will remain a residual asymmetry.  
To estimate the asymmetry we evaluate \eref{eq:solution_2} with $n_i > 0$ and subtract the same function with $n_i < 0$ to find
\begin{align}
	\frac{n_{\BpL}}{n_{i}} \approx 2 \frac{G_2 n_{i}}{\Gamma_{\BpL}} \, e^{- \Gamma_{\BpL} \Delta t} \ ,
\end{align}
where $\Delta t$ is the time elapsed between the creation of the $(B+L)$ asymmetry $n_{i}$ and the completion of the phase transition.  
To achieve a relic baryon asymmetry of the correct order, $\Delta t$ should not be much larger than $\Gamma_{\BpL}^{-1}$, in order to avoid erasing too much of the asymmetry.  In the most optimistic case, $n_{i} = n_{\gamma}$, we find $n_{\BpL} / n_{\gamma} \sim 10^{-10}$ for 
\begin{align}
	\Gamma_{\BpL} \Delta t \sim 24 + \log \frac{G_2 n_{\gamma}}{\Gamma_{\BpL}} \per
\end{align}
It is worth noting, however, that the exponential sensitivity of this result implies that a high degree of parametric tuning is required to ensure that the predicted baryon asymmetry is neither too large nor too small.  

\section{Summary and Discussion}\label{sec:Conclusion}

Although the electroweak theta term can be removed by a field redefinition in the SM, the vacuum angle contributes to a physical, invariant phase $\thetaEW$ in some minimal extensions of the SM.  
In this case, $\thetaEW \neq 0$ can provide a new source of CP-violation, acting in an analogous manner to the better-known QCD vacuum angle in the strong interactions.  
Effects of the theta term at zero temperature entail an instanton suppression, which renders them effectively unobservable, and implies that the CP-violating phase $\thetaEW$ is experimentally unconstrained.      
At finite temperature, however, non-perturbative effects become unsuppressed, and in fact the electroweak sphaleron plays a key role in many models of baryogenesis.  
In this paper we have initiated a study of how CP-violation in the sphaleron sector, arising from the electroweak vacuum angle, might impact baryogenesis.  

In \sref{sec:CPV_Finite} we have argued that CP-violation due to the electroweak vacuum angle will lead to an asymmetric erasure of an initial $(B+L)$ asymmetry.  
The central point is that an initial excess of antibaryons will reach equilibrium more quickly than one of baryons, or vice versa, depending on the sign of the CP-violating phase $\thetaEW$, as shown in \eref{eq:solution} and \fref{fig:solns}.  
In \sref{sec:Baryogenesis} we have considered the implications for models of baryogenesis that rely on sphalerons to provide B-number violation.  
Since the CP-violation shows up as a higher order effect in the kinetic equation (it is proportional to the square of the charge density; see \eref{eq:kin_eqn_extended}), it appears to have a negligible impact on models of baryogenesis in which the charge asymmetry is only $\sim 10^{-10}$ of the entropy density.  
In models with a larger initial asymmetry, the CP-violating behavior of the sphaleron can play a nontrivial role in baryogenesis.  
Ideally one would like to construct a model of baryogenesis using only the CP-violation associated with the EW sphaleron.  

Our research opens a number of avenues for potentially interesting future work.  
First, a more complete understanding of the mechanisms we have discussed could be obtained through a rigorous calculation of the quantity $G_2$, involving a fully quantum mechanical treatment of the interference effects at finite temperature.  
Second, beyond the situation discussed here, it is worth noting that the QCD sector also has a process that mediates the anomalous violation of chiral charge at finite temperature; this is the so called strong sphaleron \cite{McLerran:1990de}.  
It would be interesting to investigate how the CP-asymmetric charge erasure that we discussed in \sref{sec:CPV_Finite} manifests in the strong sector.  
In light of the strong constraints on $\bar{\theta}_{\QCD}$ from bounds on the neutron EDM, we do not anticipate quantitatively significant effects in the SM.  
However, if the vacuum angle is elevated to an axion field, then it could have taken a larger value in the early universe, and thus the QCD and EW axions (see, {e.g.} \cite{McLerran:2014daa}) present a third direction for future work.

\quad \\
\noindent
{\bf Acknowledgments:} 
We are grateful to a number of people for discussions, including Daniel Chung, Larry McLerran, Edward Shuryak, and Tanmay Vachaspati.  
A.J.L. was supported by the National Science Foundation under grant number PHY-1205745 and the Department of Energy under Grant No.\ DE-SC0008016.
The work of M.T. was supported in part by US Department of Energy (HEP) Award DE-SC0013528.

\appendix

\section{Derivation of the Kinetic Equation}\label{app:derive_G2}

In this appendix we derive the kinetic equation for the $(B+L)$ charge density, given by \eref{eq:kin_eqn_extended}.  
The leading order term was first worked out in \rref{Khlebnikov:1988sr} using the Zubarev density matrix \cite{Zubarev:1974} (see also \cite{Mottola:1990bz}).  
We follow the same approach here, and extend it by calculating and retaining the subleading term.  In this section we drop the ``B+L'' subscript on the charge current and density for clarity of presentation. 
\renewcommand{\BpL}{}

When the theory contains an explicit $(B+L)$-violating operator $\Ocal(x)$, the anomaly equation (\ref{eq:anomaly_equation}), is extended to the current conservation equation 
\begin{align}\label{app:CCE}
	\partial_{\mu} j^{\mu}_{\BpL} = 2\frac{N_g g^2}{32\pi^2}W_{\mu \nu}^{a} \widetilde{W}^{a \, \mu \nu} + \delta\Ocal \ ,
\end{align}
where $\delta \Ocal$ is the Noether variation of $\Ocal$.    
Taking the expectation value and performing a spatial average gives the initial form of the kinetic equation 
\begin{align}\label{app:KE}
	\frac{\partial n_{\BpL}}{\partial t} = 2 N_{g} \langle \frac{g^2}{32\pi^2}\overline{W_{\mu \nu}^{a} \widetilde{W}^{a \, \mu \nu}} \rangle + \delta \Ocal \ ,
\end{align}
where $n_{\BpL} = \langle \overline{j_{\BpL}^{0}} \rangle$ is the charge density, and the bar denotes spatial averaging.  
In thermal equilibrium the density matrix is simply
\begin{align}\label{eq:rho0_def}
	\rho_{0} = \frac{e^{-\beta H}}{{\rm Tr} \, e^{- \beta H}} \com
\end{align}
and $n_{\BpL}$ is static, since the sources vanish.  
When the system is prepared with an initial asymmetry, its return to equilibrium is described by the Zubarev density matrix \cite{Zubarev:1974}
\begin{align}\label{eq:rhoZub_def}
	\rho_{\tinyZ}(t) = \frac{e^{-\beta (H + h(t))}}{{\rm Tr} \, e^{-\beta (H + h(t))}}  \ ,
\end{align}
where
\begin{align}\label{eq:h_def}
	h(t) \equiv - \lim_{\varepsilon \to 0} \int_{-\infty}^{t} dt^{\prime} \, \varepsilon \, e^{\varepsilon (t^{\prime}-t)} \mu(t^{\prime}) N(t^{\prime}) \per
\end{align}
The charge operator is $N_{\BpL}(t) = \int \! d^3x \, j_{\BpL}^{0}(x)$, and the chemical potential $\mu(t)$ parameterizes the charge asymmetry via
\begin{align}\label{eq:n_to_mu}
	n_{\BpL}(t) = c_{\rm dof} \, N_{g} \, \mu(t) T^2 \com
\end{align}
where $c_{\rm dof}$ is the effective number of fermionic degrees of freedom per generation in equilibrium with the sphaleron.  

Assuming that $\mu(t)$ is slowly varying, we approximate 
\begin{align}\label{eq:h_approx}
	h(t) 
	\approx - \mu(t) N(t)
	+ \mu(t) \lim_{\varepsilon \to 0} \int_{-\infty}^{t} \! \! \! dt^{\prime} \, 
	e^{\varepsilon (t^{\prime}-t)} \dot{N}(t^{\prime}) \ .
\end{align}
Furthermore, when the charge asymmetry is small, we can expand $\rho_{\tinyZ}$ for $h \ll H$ to find 
\begin{align}\label{eq:rhoZub_final}
	\rho_{\tinyZ} \approx \Bigl( 1 + C_1 - \langle C_1 \rangle_0 + C_2 - \langle C_2 \rangle_0 \Bigr) \rho_0  \ ,
\end{align}
where $\langle \cdot \rangle_{0}$ is an expectation value with respect to $\rho_{0}$, and we have introduced the operators 
\begin{align}
	& C_1 \equiv \ - \beta \int_{0}^{1} \! d\lambda \, e^{- \lambda \beta H} \, h \, e^{\lambda \beta H} \label{eq:C1_def} 
	\\
	& C_2 \equiv 
	\ \frac{\beta^2}{2} \int_{0}^{1} \! d\lambda \! \int_{0}^{1} \! d\sigma \, 
	\Bigl[ 
	(\lambda) e^{-\sigma \lambda \beta H} \, h \, e^{-(1-\sigma) \lambda \beta H} \, h \, e^{\lambda \beta H} \nn
	& \quad \qquad + (1-\lambda) \, e^{-\lambda \beta H} \, h \, e^{-\sigma (1-\lambda) \beta H} \, h \, e^{(\sigma + \lambda - \sigma \lambda) \beta H}
	\Bigr] \label{eq:C2_def} \per
\end{align}
If $h$ and $H$ commute with each other, then we have $C_1 = - \beta h$ and $C_2 = \beta^2 h^2 / 2$.  

The left side of the kinetic equation (\ref{app:KE}), can be written as $\dot{n}_{\BpL} = \langle \dot{N}_{\BpL} \rangle / V$.  
Using (\ref{eq:rhoZub_final}), the expectation value of the number operator is
\begin{align}
	\langle \dot{N} \rangle 
	& \approx \langle \dot{N} \rangle_0 + \langle \dot{N} C_1 \rangle_0 - \langle \dot{N} \rangle_0 \langle C_1 \rangle_0 
	+ \langle \dot{N} C_2 \rangle_0 - \langle \dot{N} \rangle_0 \langle C_2 \rangle_0 \ .
\end{align}
There is no spontaneous charge generation in equilibrium, $\langle \dot{N} \rangle_0 = 0$, and we find 
\begin{align}
	\langle \dot{N} \rangle 
	\approx \langle \dot{N} C_1 \rangle_0 + \langle \dot{N} C_2 \rangle_0 
\end{align}
up to higher order terms in $\beta \mu \ll 1$, which we have dropped.  
The result $\langle \dot{N} \rangle_0 = 0$ follows from the time-reversal invariance of $\rho_{0}$.  

Using (\ref{eq:h_approx}) and (\ref{eq:C1_def}), the leading term can be written as 
\begin{align}
	\langle \dot{N} C_1 \rangle_0 = & - \mu(t) \beta \lim_{\varepsilon \to 0} \int_{-\infty}^{t} dt^{\prime} e^{\varepsilon (t^{\prime} - t)} 
	\int_{0}^{1} \! d\lambda \left< \dot{N}(t) e^{-\lambda \beta H} \dot{N}(t^{\prime}) e^{\lambda \beta H} \right>_0 \ ,
\end{align}
where, in simplifying, we have used $\langle \dot{N} N \rangle_0 = 0$, which follows from the CPT invariance of $\rho_0$.  
Now, using (\ref{eq:n_to_mu}) we obtain
\begin{align}
	\langle \dot{N} C_1 \rangle_0
	\approx -\GammaB V n_{\BpL}(t) \ ,
\end{align}
where we have defined  
\begin{align}\label{eq:GammaB_def}
	\GammaB \equiv \, 
	& \frac{1}{c_{\rm dof} N_g T^3 V} \lim_{\varepsilon \to 0} \int_{-\infty}^{t} dt^{\prime} e^{\varepsilon (t^{\prime} - t)}  
	\int_{0}^{1} d\lambda \, \left< \dot{N}(t) \, e^{-\lambda \beta H} \dot{N}(t^{\prime}) e^{\lambda \beta H} \right>_0 
\com
\end{align}
which is the rate of $(B+L)$ violation.  
Using the current conservation equation, (\ref{app:CCE}), it is a standard calculation to relate this matrix element to the Chern-Simons number diffusion coefficient $\gamma_{\rm diff}$.  
One finds $\GammaB \sim N_{g} \gamma_{\rm diff} / c_{\rm dof} T^3$, plus a term that arises from the explicit $(B+L)$-violating operator $\Ocal(x)$.  
We have dropped an inconsequential $O(1)$ coefficient (see \cite{Mottola:1990bz} for more details.)  

We evaluate the quadratic term $\langle \dot{N} C_2 \rangle_0$ similarly.  First, we substitute $h$ from (\ref{eq:h_approx}) into $C_{2}$ from (\ref{eq:C2_def}), and use (\ref{eq:n_to_mu}) to eliminate $\mu(t)$ for $n(t)$.  
This gives 
\begin{align}
	\langle \dot{N} C_2 \rangle_0 \approx G_{2} V n_{\BpL}(t)^2
	\com
\end{align}
where 
%
\begin{align}\label{eq:G2_appendix}
	G_2 \equiv 
	& - \frac{1}{2 c_{\rm dof}^2 N_g^2 V T^6} \Biggl\{
	\int_{0}^{1} \! d\lambda \int_{0}^{1} \! d\sigma \Biggl[ 
	(\lambda) \left< \dot{N}(t) e^{-\sigma \lambda \beta H} N(t) \, e^{-(1-\sigma) \lambda \beta H} N(t) \, e^{\lambda \beta H} \right>_{0} \\
	& \qquad + (1-\lambda) \left< \dot{N}(t) \, e^{-\lambda \beta H} \, N(t) \, e^{-\sigma (1-\lambda) \beta H} \, N(t) \, e^{(\sigma + \lambda - \sigma \lambda) \beta H} \right>_{0} \Biggr] \nn
	& \ - \int_{0}^{1} \! d\lambda \, \int_{0}^{1} \! d\sigma \, \lim_{\varepsilon \to 0} \int_{-\infty}^{t} dt^{\prime} \, e^{\varepsilon(t^{\prime}-t)}
	\Biggl[
	(\lambda) \, \left< \dot{N}(t) e^{-\sigma \lambda \beta H} \, \dot{N}(t^{\prime}) \, e^{-(1-\sigma) \lambda \beta H} \, N(t) \, e^{\lambda \beta H} \right>_{0} \nn
	& \qquad + (\lambda) \, \left< \dot{N}(t) e^{-\sigma \lambda \beta H} \, N(t) \, e^{-(1-\sigma) \lambda \beta H} \, \dot{N}(t^{\prime}) \, e^{\lambda \beta H} \right>_{0} \nn
	& \qquad + (1-\lambda) \, \left< \dot{N}(t) e^{-\lambda \beta H} \, N(t) \, e^{-\sigma (1-\lambda) \beta H} \, \dot{N}(t^{\prime}) \, e^{(\sigma + \lambda - \sigma \lambda) \beta H} \right>_{0} \nn
	& \qquad + (1-\lambda) \, \left< \dot{N}(t) e^{-\lambda \beta H} \, \dot{N}(t^{\prime}) \, e^{-\sigma (1-\lambda) \beta H} \, N(t) \, e^{(\sigma + \lambda - \sigma \lambda) \beta H} \right>_{0}
	\Biggr] \nn
	& + \int_{0}^{1} \! d\lambda \int_{0}^{1} \! d\sigma \lim_{\varepsilon \to 0} \int_{-\infty}^{t} dt^{\prime} \lim_{\varepsilon^{\prime} \to 0} \int_{-\infty}^{t} dt^{\prime \prime} e^{\varepsilon (t^{\prime} - t)} e^{\varepsilon^{\prime}(t^{\prime \prime} - t)} \Biggl[ 
	(\lambda) \left< \dot{N}(t) e^{-\sigma \lambda \beta H} \dot{N}(t^{\prime}) \, e^{-(1-\sigma) \lambda \beta H} \dot{N}(t^{\prime \prime}) \, e^{\lambda \beta H}  \right>_{0} \nn
	& \qquad + (1-\lambda) \left< \dot{N}(t) \, e^{-\lambda \beta H} \, N(t^{\prime}) \, e^{-\sigma (1-\lambda) \beta H} \, N(t^{\prime \prime}) \, e^{(\sigma + \lambda - \sigma \lambda) \beta H} \right>_{0}
	\Biggr\} \nonumber \ .
\end{align}
Unlike in the previous calculation, it is not obvious that we can use CPT invariance to drop any of these terms.  
Na\"{\i}vely, operator products like $\dot{N} \dot{N} N $ are odd under CPT, but the standard analysis is made more complicated because the operators are evaluated at different times.  
However, it is important to note that these triple operator products are odd under CP, which takes $N \to - N$.  
Therefore, CP-invariance implies $G_{2}= 0$, and we expect $\langle \dot{N} C_2 \rangle_0$ to be nonzero when the system is CP-violating.  
Thus, the kinetic equation for the $(B+L)$ charge density takes the form (\ref{eq:kin_eqn_extended}).

\bibliographystyle{h-physrev5}
\bibliography{refs--EW_Vac_Angle}

\end{document}